\pdfoutput=1

\documentclass[11pt]{article}

\usepackage[final]{naacl2021}

\usepackage{times}
\usepackage{latexsym}

\usepackage[T1]{fontenc}

\usepackage[utf8]{inputenc}

\usepackage{microtype}

\title{Steve: LLM Powered ChatBot for Career Progression }
\author{Author 1 \and ... \and Author n \\
         Address line \\ ... \\ Address line}

\author{Naveen Mathews Renji, Balaji Rao, Carlo Lipizzi \\
Stevens Institute of Technology \\
\{n3, brao, clipizzi\}@stevens.edu}

\begin{document}
\maketitle
\begin{abstract}
The advancements in systems deploying large language models (LLMs), as well as improvements in their ability to act as agents with predefined templates, provide an opportunity to conduct qualitative, individualized assessments, creating a bridge between qualitative and quantitative methods for candidates seeking career progression. In this paper, we develop a platform that allows candidates to run AI-led interviews to assess their current career stage and curate coursework to enable progression to the next level. Our approach incorporates predefined career trajectories, associated skills, and a method to recommend the best resources for gaining the necessary skills for advancement. We employ OpenAI API calls along with expertly compiled chat templates to assess candidate competence. Our platform is highly configurable due to the modularity of the development, is easy to deploy and use, and available as a web interface where the only requirement is candidate resumes in PDF format. We demonstrate a use-case centered on software engineering and intend to extend this platform to be domain-agnostic, requiring only regular updates to chat templates as industries evolve. 
\end{abstract}

\section{Introduction}
Career progression is a process requiring a blend of intentional efforts from candidates, as well as subjective insights from educational support systems based on industry-driven needs. Traditional methods often rely on Human Resource (HR) professionals to build protocols to assist in career progression. These are one-dimensional tools and fail to capture the full spectrum of a candidate's potential and expertise. Our work introduces a comprehensive system that combines AI models and structured interview protocols to deliver detailed, actionable assessments and concrete suggestions in the form of certifications to acquire the required skills.

\begin{figure}[h!]
\centering
    \includegraphics[width=0.49\textwidth]{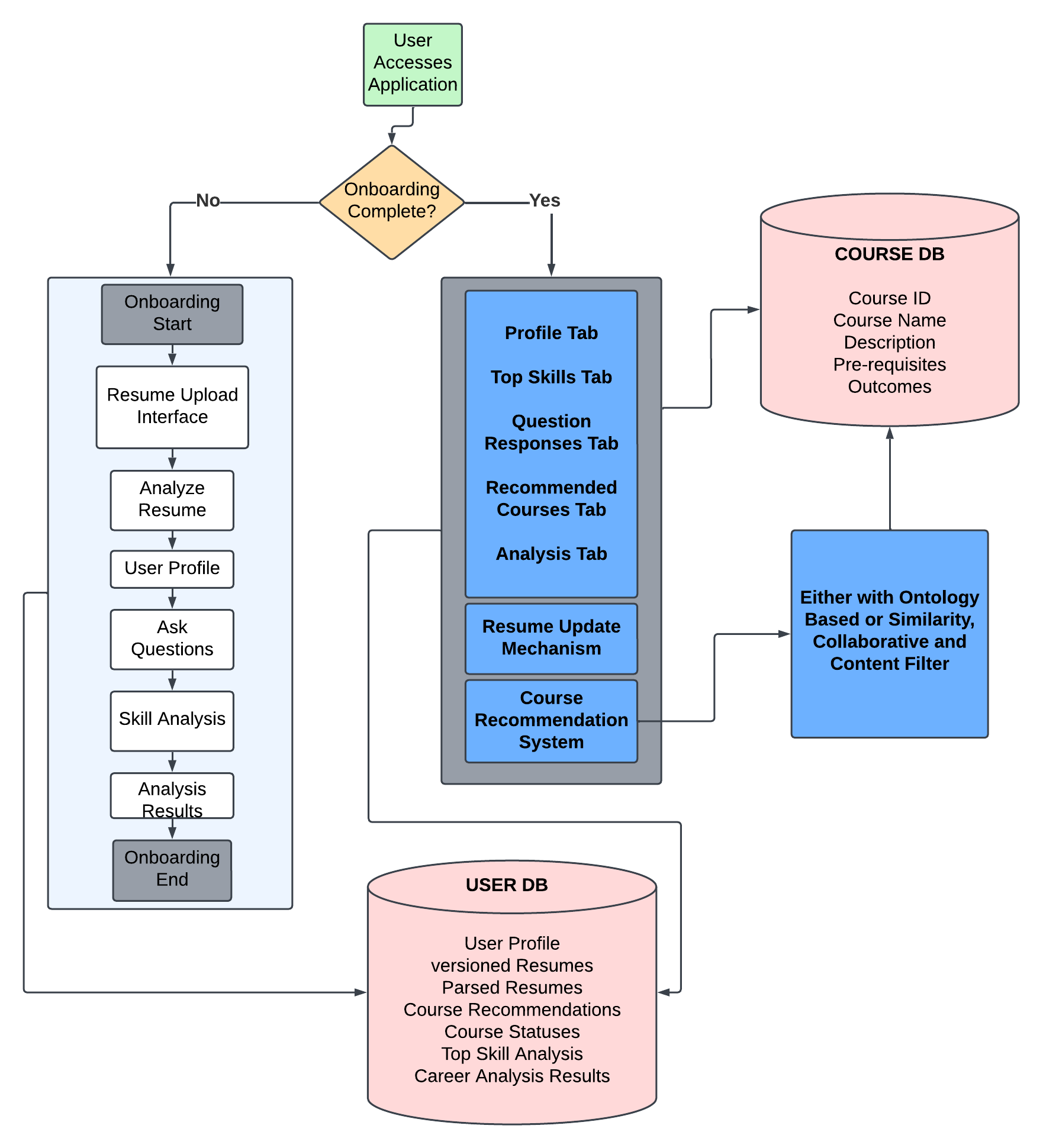}
    \caption{AI Powered Career Coach}
    \label{fig1}
\end{figure}

Key contributions of our work include:
\begin{itemize}
    \item A modular system for parsing resumes, assessing skills, recommending career paths, and courses available on popular platforms.
    \item Interactive AI conversational agents tailored to individual profiles.
    \item A novel methodology that integrates qualitative and quantitative career progression insights.
\end{itemize}

\section{Related Work}
The integration of artificial intelligence (AI) and natural language processing (NLP) into most professional development systems has redefined the bar for the employee experience. One main area identified is in personalized learning and coaching systems. Career coaching is a powerful way to enhance employee productivity and strengthen succession pipelines for businesses. AI Systems in a learning context help accelerate skill development at the level of the individual, and they can optimize learning at the level of the organization, leading to personal progression as well as organizational progress~\cite{guenole2018business}. These systems have demonstrated their potential in transforming career counseling by offering scalable and personalized solutions. Shilaskar et al. present a dual-layered system that combines machine learning-based career predictions with a RASA framework chatbot to deliver tailored advice to students~\cite{shilaskar2024conversational}. In the same vein, Lee et al. discuss a chatbot-driven system for providing informational support, career planning interventions, and personalized feedback~\cite{lee2019intelligent}.

NLP systems are commonly used as resume parsing tools for organizations mainly due to their ability to understand and
parse the unstructured written language and context-based information~\cite{sinha2021resume}. There are systems like the one proposed by Kothari et al. that use ChatGPT-based methods to assess skill gaps~\cite{kothari2024interviewease}. We adopt a similar approach in our prototype to extract structured data such as education, skills, and inferred soft skills from resumes.

The use of hierarchical career trees and semantic-enhanced transformers has significantly impacted career path recommendations. Nguyen et al. discuss an approach for extracting job hierarchies~\cite{xu2018extracting} where they use the Difficulty of Promotion (DOP) from one job
title to another based on the assumption that
a longer tenure usually implies a greater difficulty
to be promoted. While Guan et al. present a skill-aware recommendation model based on semantic-enhanced transformers to parse job descriptions and associate profiles with missing candidate qualifications~\cite{guan2024jobformer}.

With the rise in personalized AI chatbots like ChatGPT~\cite{ChatGPT}, many platforms have been developed that concentrate on personalized learning pathways and skill development systems. There is an emphasis on using ontologies to guide such planning. Chen et al. emphasize how structured knowledge representation can guide dynamic educational plans~\cite{chen2009ontology}. Their approach has influenced how systems recommend curated courses and learning projects to address skill gaps. In a similar way Li et al. provide a detailed gap analysis, emphasizing the alignment of educational pathways with industry requirements~\cite{li2021data}. 

Mentorship plays a pivotal role in career development. There is a focus on the need for a individual-centered mentoring models to assist in career development, to ensure effective strategies for learning and progression~\cite{montgomery2017mapping}. Furthermore, the conversational capabilities of career coaching systems that utilize AI is highlighted by Lee et al.~\cite{lee2019intelligent}. 

These works are among many others that highlight the evolution of AI and NLP techniques in transforming career guidance systems. Our system attempts to bridge the gap between career development, assessment, and learning systems by integrating hierarchical career trees, semantic search, and conversational AI, by addressing limitations in traditional approaches, providing personalized, scalable career coaching.

\section{Methodology}

Our system is built using NLP techniques and AI methodologies for the technical stack and the React tech stack for the user interface components. We leverage structured and unstructured data to provide actionable insights.

\subsection{Resume Parsing}

The candidates upload their resumes on the portal. The uploaded resumes are preprocessed by extracting text from uploaded files using PyPDF2. We tokenize the text and split it into manageable chunks to accommodate model token limits. We use Function Calling to generate structured outputs to ensure the model's output exactly matches our specified JSON schema.
We use Named Entity Recognition within the
GPT model to extract structured information such as names, contact details, education, and experience.
We prompt the model to infer soft skills from contextual cues within the text, considering roles and accomplishments. Once the chunks are processed independently, they are merged and duplicates are removed while maintaining unique entries.

\subsection{Career Path Analysis}
We define a career tree in JSON that outlines hierarchical role progressions within the industry. This predefined tree is highly configurable and serves as the backbone for analyzing and suggesting career advancements. Using NLP-based analysis of the extracted resume's content, we can identify the candidate's current position and recommend potential short-term and long-term career growth opportunities based on hierarchical role relationships. The structured nature of the career hierarchy allows us to associate each node in the career tree with a specific role.

This career tree is included in the OpenAI API prompt to ensure that the model evaluates the candidate's resume within the context of defined roles and growth pathways. The candidate's current role is mapped to a node in the career tree using semantic similarity. We evaluate the context and responsibilities described in the resume to identify a match. Recommendations are tailored based on user-provided data, such as skills and experience extracted from the resume, as well as answers to specific questions.

Once the current position is identified, the system uses the hierarchical structure of the career tree to provide tailored recommendations for career progression. Roles in the next-positions attribute of the identified node are suggested as immediate growth opportunities. Roles in the second-jump-positions attribute are recommended as advanced positions achievable with further development.

\subsection{Skill Assessment}
Our system matches the candidate's extracted skills (technical and soft) against a set of predefined required skills for their current role. These requirements are stored in a structured JSON file (skills.json) and include detailed descriptions and expectations for various positions. This is also included in the prompt to ensure that the model is able to assess each candidate against the desired skills. Each of these skills is categorized into one of four proficiency levels: beginner, intermediate, advanced, or expert. These levels are determined based on factors such as the duration of experience with the skill, the implicit or explicit frequency of use in past roles or projects, and mentions in the candidate's resume or additional Q\&A inputs.

By incorporating additional candidate-provided details during the Q\&A portion of the assessment, our system refines the skill analysis to ensure relevance and alignment with the candidate's aspirations. This evaluation highlights gaps by identifying skills that are required for the candidate's next role but are either absent or insufficiently developed. 

\subsection{Interactive Coaching}
The skill assessment and the gaps identified allow us to tailor a development plan that aligns with the candidate’s current role and future career goals. 
This includes skill enhancement strategies and curated learning resources. These resources are compiled by the organization deploying our system, this helps with the alignment of organizational needs and the candidate's aspirations. 

Our NLP system delivers a conversational and adaptive experience, guiding candidates through their professional development journey. These prompts are dynamically generated using parsed resume data, career path analysis, and Q\&A responses. The prompts are tailored to address the candidate's unique strengths, weaknesses, and career trajectory. Our system is sensitive to the candidate's most recent experiences, ensuring relevance and continuity in the conversation. This process helps compile a summary of key takeaways, including strengths, specific skill gaps, and improvement areas identified during the analysis. It concludes with motivational insights about the candidate and suggests next steps for career development.

\subsection{Qdrant Tool}
To ensure the relevance and personalization of each individual skill development recommendation, we use a vector database, Qdrant~\cite{qdrant}, with a pretrained HuggingFace embedding model (all-MiniLM-L6-v2) for semantic search and retrieval. Course details, which include the title, description, URL, and outcomes, are ingested from structured CSV files and embedded into high-dimensional vectors. Each course is represented by a composite vector generated from its title, description, and learning outcomes, ensuring that the embedded representation captures the semantics of the course content. These vectors are stored in Qdrant, which organizes the data into collections for efficient similarity-based retrieval. We use cosine similarity as the distance metric to recommend top-matching courses, which are identified in the gaps during the skill assessment.

\section{System Design}
Our system is intended to be used by a large number of users. To ensure scalability as well as asynchronous operations and to ensure non-blocking API calls for tasks such as resume parsing and database updates we use FastAPI~\cite{2018fastapi}. To ensure our system is ingesting resumes within the token limit we use the Tiktoken~\cite{tiktoken} as the tokenizer to processes text in manageable chunks. We use MongoDB~\cite{MongoDB} as the database to store the user profiles. This is used for indexing and making database queries where the user journey and resumes are stored.

\subsection{Career Path Curation}
The current stage of our system relies heavily on a subject matter expert (SME) to design and configure career trajectories and the corresponding skill requirements for each role, defined in JSON format. This approach ensures that the system remains adaptable and tailored to industry-specific standards and also allows customizations for organization-specific requirements to align recommendations with role objectives. Our career trajectory framework leverages a predefined hierarchical structure, outlining potential growth opportunities across various roles.

\subsection{Course Database Curation}
For the purpose of this prototype, we concentrated on a case study centered mainly around Software Engineering, Machine Learning, Artificial Intelligence, and Software Development. We used a script that scraped and extracted course-related information from Coursera's platform. We utilized a script that used CSS selectors to extract critical details such as course titles, skills, and course descriptions. From this web scraping and data extraction script, we obtained a structured and comprehensive dataset of Coursera courses for academic exploration.

From this dataset, we filtered for more specific keywords from the skills column that included words like : Python, TensorFlow, Agile, Git, DevOps, SQL, R and so on. We then used OpenAI API calls to analyze each course description and generate a set of measurable, industry-aligned course outcomes. This was used as our course database that was ingested into the Qdrant Vector Database.

\section{Implementation Details}

\subsection{Technologies}
We utilize a robust technology stack to deliver scalable, efficient, and personalized user experience with the career recommendation system.

\subsubsection{Frontend Stack}
Our main user interface is built with React.js for resume uploads, Q\&A sessions, and personalized recommendations. The frontend communicates with backend APIs to display dynamic results like career path analyses, skill assessment outcomes, and many other functionalities to help with career progression.

\subsubsection{Backend Stack}
We utilize FastAPI for asynchronous processing and other high-performance API calls. It handles operations like file parsing, career path analysis, and database interactions.
We will be hosting our system on the Microsoft Azure cloud platform to address scalability and concurrency bottlenecks.

\subsubsection{Database Systems}
Our system utilizes 2 main databases. One for more classical forms of data storage and another vector database for NLP applications.

\begin{enumerate}
    \item MongoDB. It serves as our primary database for storing structured data, including parsed resumes, skill analysis results, Q\&A responses, and recommendations. Its document-based architecture integrates with JSON-like data structures.
    \item Qdrant Vector Database. We use this to assist in semantic search and recommendation tasks. This platform stores vector embeddings of our course descriptions and learning outcomes, for similarity-based retrieval of course recommendations.
\end{enumerate}

\subsubsection{Natural Language Processing Tools}
At the current stage of our prototype, we use OpenAI’s GPT models extensively for extracting data, analyzing career paths, identifying skill gaps, and generating recommendations through structured API calls. The system utilizes Function Calling to ensure outputs conform to our predefined schemas for consistency and reliability.

\subsection{Data Flow and Integration}
Candidates upload their resumes through a dedicated API endpoint, which is a webpage. The uploaded PDF Files are saved temporarily, parsed, and structured data extracted using the OpenAI API call, and is stored in MongoDB.

The careerPath module maps the parsed resume data to a predefined hierarchical career tree stored as a JSON structure for subsequent analysis. This career tree outlines potential progressions and growth opportunities for various roles.  We use semantic matching to map the user’s current role to the most appropriate node in the career tree. 

Role-specific questions are used by the system refine user preferences, career aspirations, and augment the skill assessment. This dynamic system ensures adaptiveness for each candidate as they progress through their career.

Our skills module evaluates the user’s technical and soft skills against benchmarks defined by the organization. Each skill is categorized into one of four proficiency levels based on implicit and explicit references in the resume as well as the Q\&A session. We flag missing or underdeveloped skills as gaps by comparing the current role and the short-term (next positions) progression. 

The entire assessment for each session along with the identified gaps is transformed into a query text the represents the gaps in the form of skills and technologies required for the next position progression. This query, generated from skill assessments, is embedded and matched against stored vectors using Qdrant’s similarity search capabilities. Based on the results of this search, courses are presented to users as personalized learning recommendations.

\section{Tool Overview}
We name our platform Steve. It is an AI-powered career coaching system, integrating natural language processing (NLP) and ontology-driven approaches to deliver personalized career guidance.

\begin{figure}[h!]
\centering
    \includegraphics[width=0.49\textwidth]{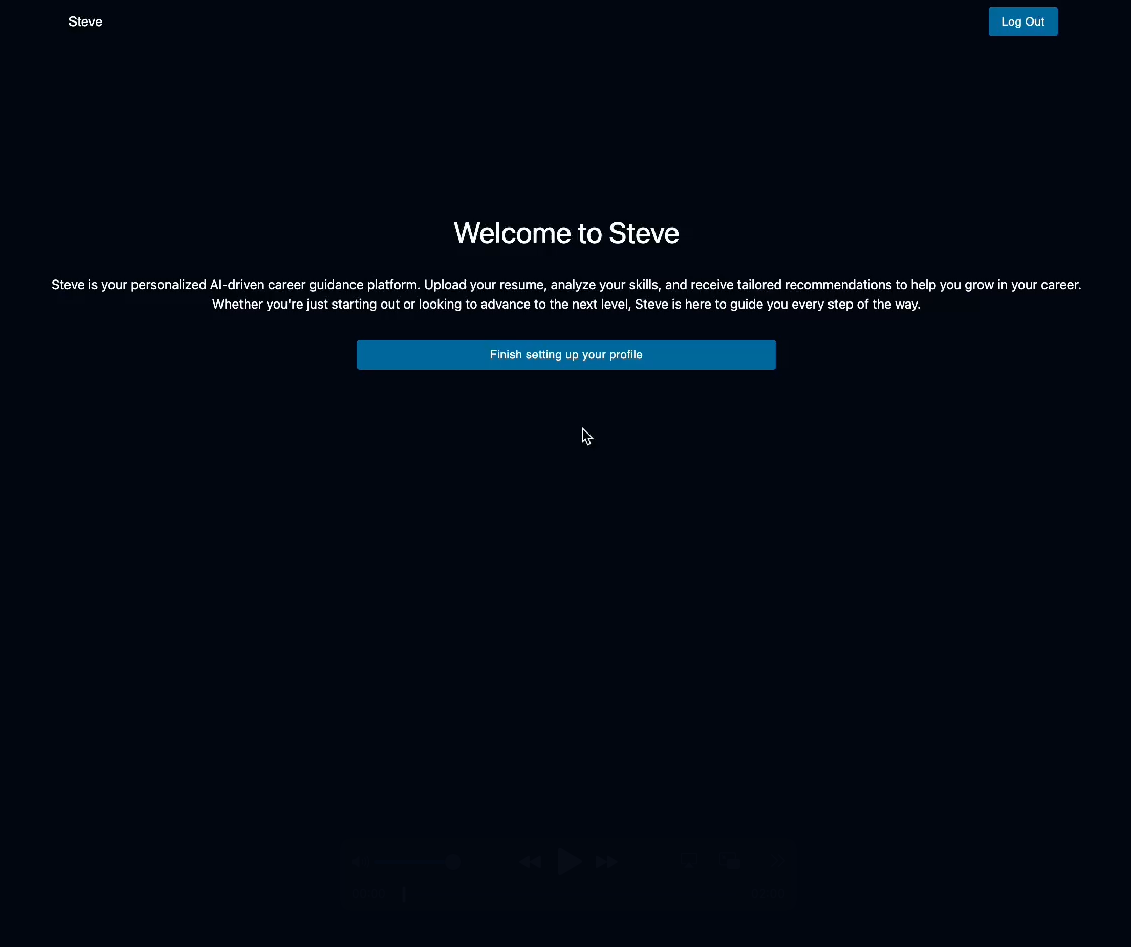}
    \caption{A screenshot of Steve's Landing Page}
    \label{fig2}
\end{figure}

The process begins with candidates setting up their profiles once they land on the intuitive interface as shown in figure \ref{fig2}. The candidates then upload their resumes and Steve processes the resume using a language model that parse unstructured information into structured data through function calls. The extracted data is organized to enable further analysis, creating a clear and comprehensive profile for each user.

\begin{figure}[h!]
\centering
    \includegraphics[width=0.49\textwidth]{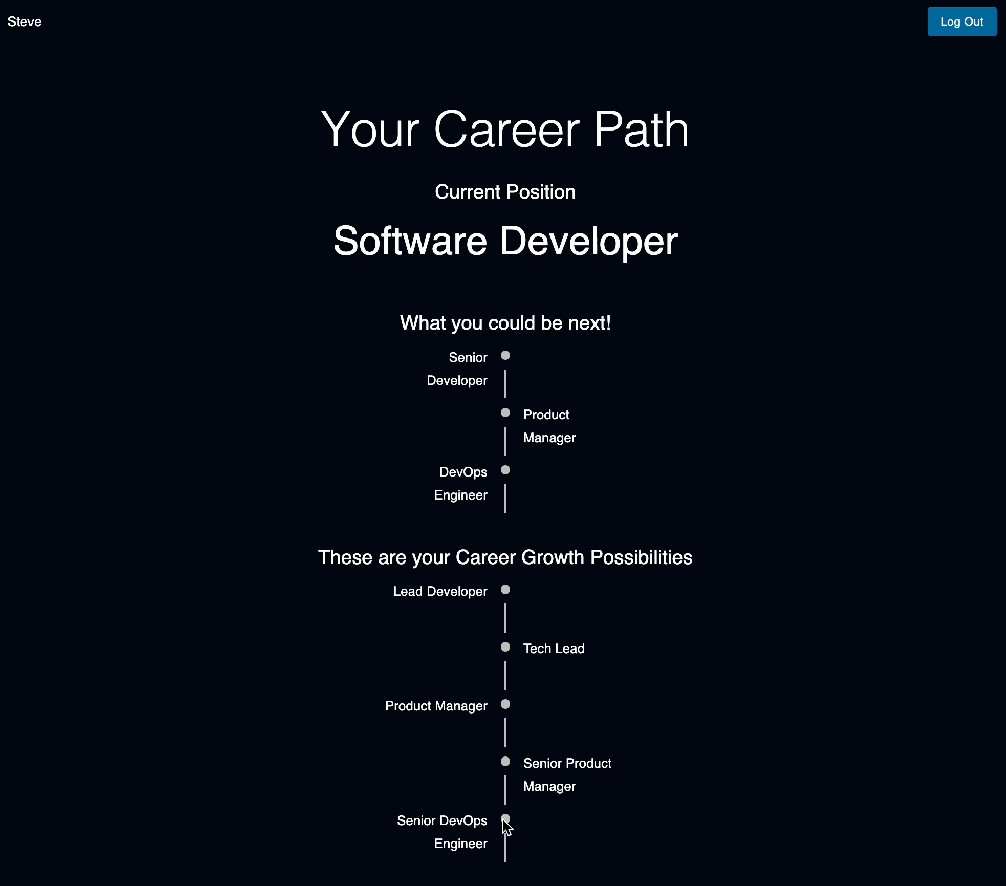}
    \caption{A screenshot of the Career path created by Steve}
    \label{fig3}
\end{figure}

Steve uses NLP techniques to extract both technical and soft skills from the user's resume and project descriptions. Steve employs an ontology-driven approach to match user profiles with predefined and configurable career paths. By analyzing the semantic similarity between the user's skills and potential career trajectories, Steve provides tailored next-possible career roles, an example of which is shown in figure \ref{fig3}.

Next, Steve interacts  with the candidate through a dynamic question-and-answer phase. Our system uses cognitive profiling to ask tailored, configurable questions that align with the user's career stage, goals, and aspirations. 

\begin{figure}[h!]
\centering
    \includegraphics[width=0.49\textwidth]{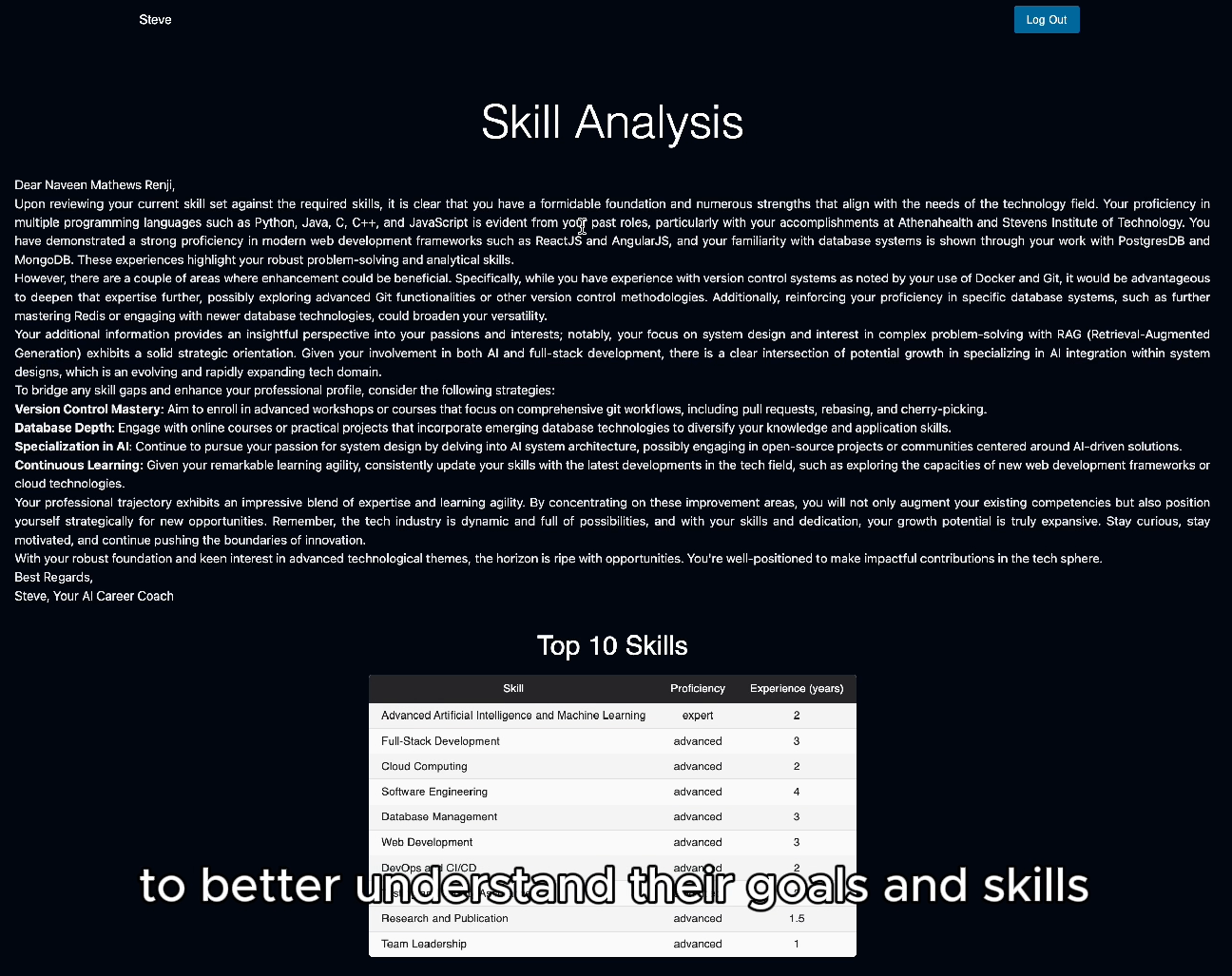}
    \caption{A screenshot of the skill report created by Steve}
    \label{fig4}
\end{figure}

To build a competency profile of the candidate, Steve compares the user's current skills derived from the resume and the Q\&A session against the requirements of their present and next roles defined in the tree. By identifying skill gaps in this way, Steve generates a detailed report that includes role-specific recommendations for improvement. An example of the skill report is shown in figure\ref{fig4}.

\begin{figure}[h!]
\centering
    \includegraphics[width=0.49\textwidth]{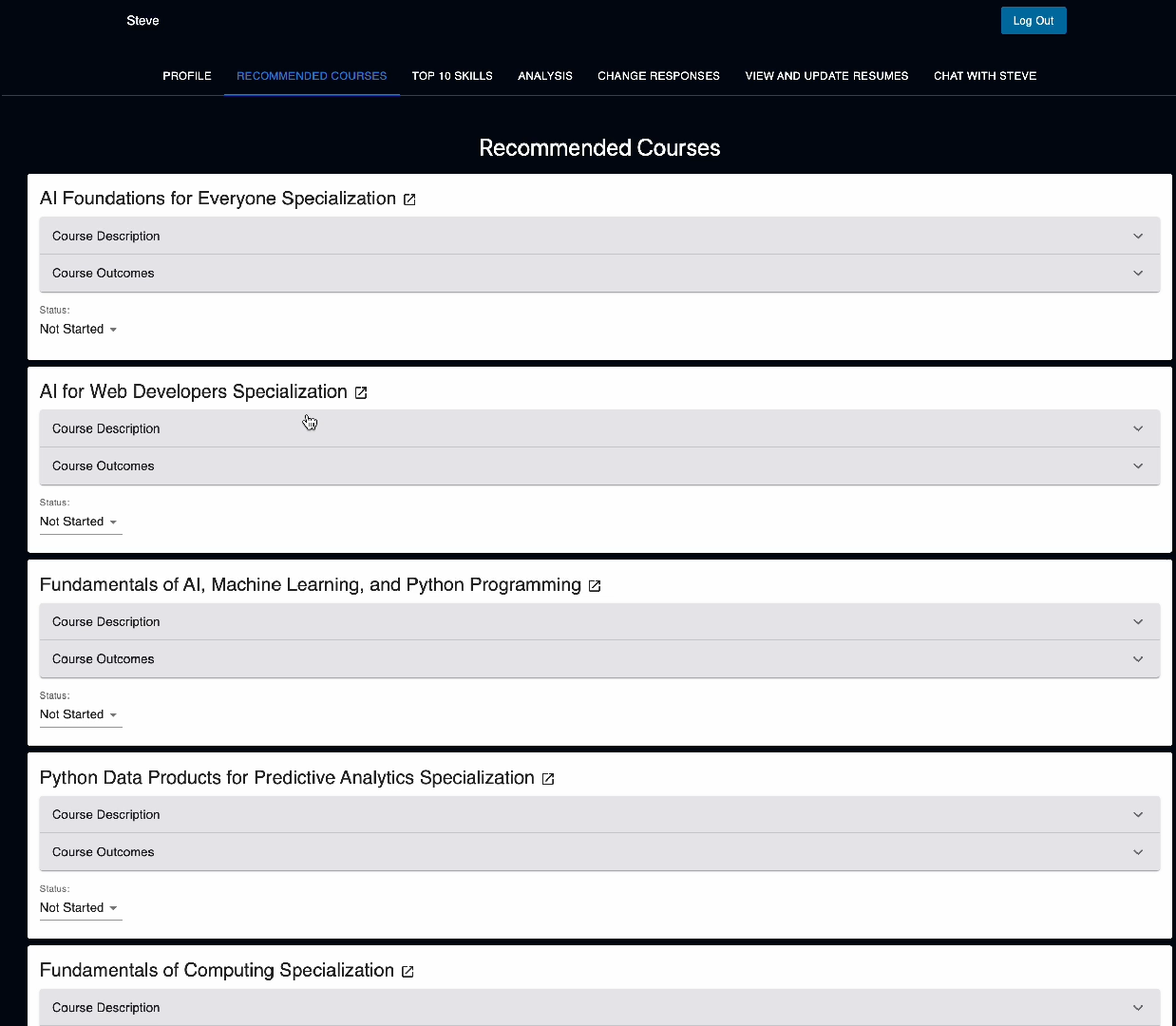}
    \caption{A screenshot of the Recommended courses}
    \label{fig5}
\end{figure}

Our system  evaluates the proficiency of the user's top 10 skills based on their duration and frequency of use and mentions in the assessments. To bridge identified skill gaps, Steve recommends targeted learning resources, such as courses from platforms like Coursera. This concludes the first interaction with Steve. This interaction builds a profile of the candidate that they can log into for their subsequent interactions, as shown in figure \ref{fig5}. 

\begin{figure}[h!]
\centering
    \includegraphics[width=0.49\textwidth]{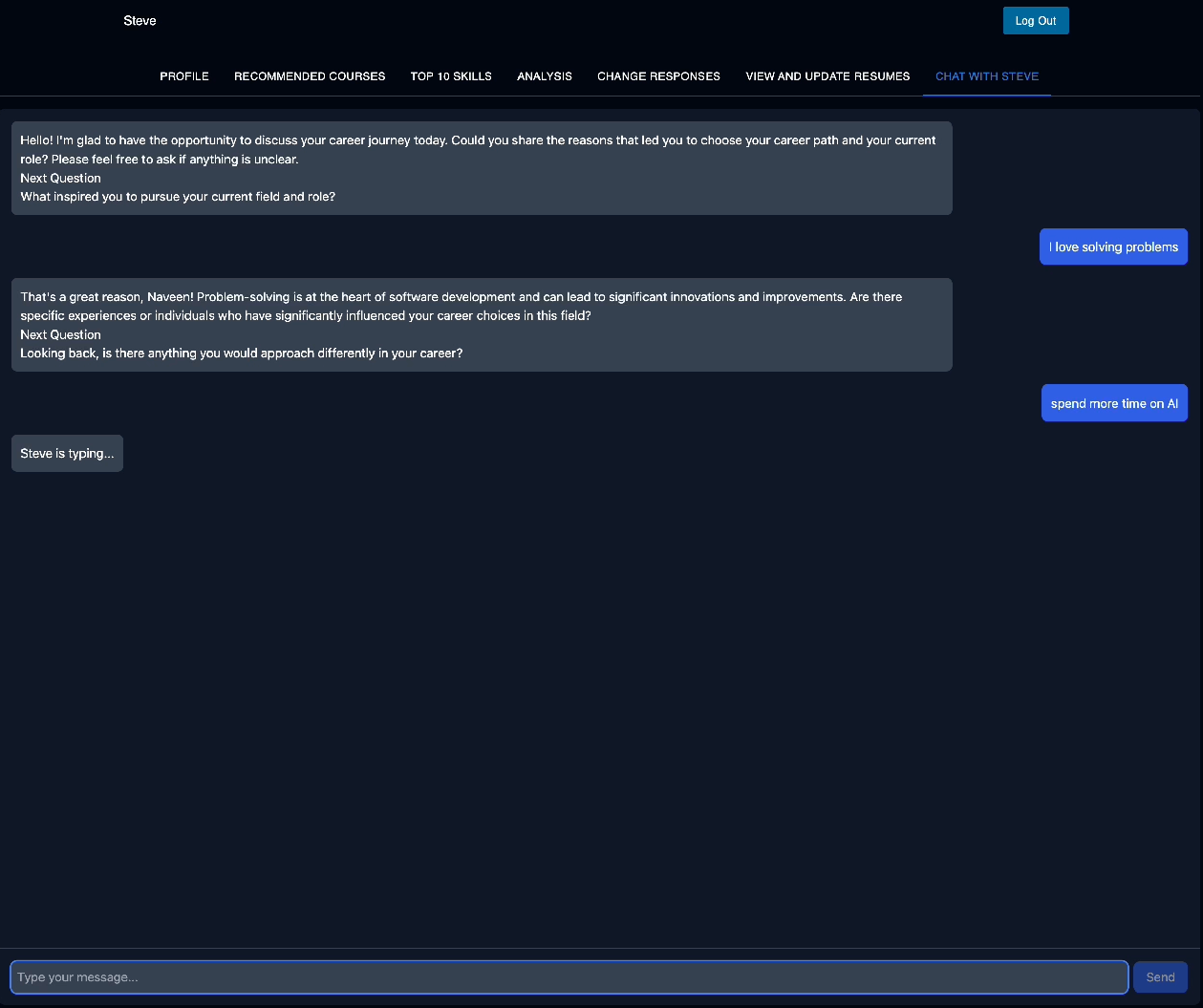}
    \caption{Chat with Steve}
    \label{fig6}
\end{figure}

Steve includes a conversational chatbot that dynamically adapts to the user's resume and skill analysis, as shown in figure \ref{fig6}. This enables personalized guidance, addressing user queries and reinforcing the insights provided by the system. Once the profile has been created, Steve keeps track of the profile's progress through a resume tracker that allows users to upload and update their resumes. This automatically recalibrates recommendations, skill analyses, and suggested courses. 

Additionally, candidates can monitor their progress with a dedicated tracker for recommended courses, ensuring accountability and continuity in their career development. The memory functionality enables candidates to revisit and revise their responses in the Q\&A phase as they progress in their careers or decide they need a pivot, which prompts real-time updates to their profile as well as the recommendations.

\section{Conclusion and Discussion}

In this work, we present Steve, an AI-powered career coaching system. We integrate advanced natural language processing (NLP), machine learning, and ontology-driven methodologies to deliver an innovative, user-centric approach towards career guidance. Steve empowers users to navigate their professional journeys with actionable insights and tailored recommendations. Our system's multifaceted pipeline enables it to  adapt to the needs and requirements of each candidate. 

The modularity of our system provides a unique opportunity to make the coach industry-agnostic. The only bottleneck is the current reliance on SMEs to curate the career hierarchies in the organization intended for use and the skills they expect the candidates in the role to be proficient in.

\section{Workforce Implications}

The rise of sophisticated chatbots  and automation technologies, exemplified by Steve, have far-reaching implications for the workforce. Industries are rapidly adopting these technologies, and the demand for highly specialized skills and continuous learning has become more pronounced. As AI Systems drive innovation, they also necessitate a rethinking of workforce development. Developing highly specialized skills and continuous learning to adapt to new AI-driven job roles is crucial\cite{rao2024anatomy}.

While systems like Steve aim to address these gaps by providing personalized, and on-demand support, ensuring that users from diverse socioeconomic backgrounds have access to high-quality career development resources, this level of advancement also raises concerns about workforce displacement and the uneven impact of automation on different sectors. We present AI tools like Steve to not only enhance individual skill development and employability, but also provide organizations with a tool for workforce planning.

\section{Ethical Considerations}
As with the use of any LLM-powered AI System, the primary concern is the potential for bias in AI based assessments, and recommendations. Most systems that purely use Large Language Models may reflect societal biases, such as gender or racial disparities in employment. However, our system tries to mitigate the propagation of this bias by using more knowledge grounded resources with the career progression and courses that are in the recommendation database.

Another ethical dimension involved is the balance between automation and human oversight. While Steve leverages AI to provide personalized recommendations, it heavily relies on subject matter experts to define career paths and skill requirements, ensuring domain knowledge and human judgment are integrated into the system at the level of its inception. There are other ethical implications such as accessibility, multilingual support and so on that might adversely affect marginalized comminutes. Such considerations remain a challenge in the ethical development and deployment of AI as a whole.

\bibliography{naacl2021}

\begin{thebibliography}{16}
\expandafter\ifx\csname natexlab\endcsname\relax\def\natexlab#1{#1}\fi

\bibitem[{Chen(2009)}]{chen2009ontology}
Chih-Ming Chen. 2009.
\newblock Ontology-based concept map for planning a personalised learning path.
\newblock \emph{British Journal of Educational Technology}, 40(6):1028--1058.

\bibitem[{Guan et~al.(2024)Guan, Yang, Yang, Zhu, Li, and Xiong}]{guan2024jobformer}
Zhihao Guan, Jia-Qi Yang, Yang Yang, Hengshu Zhu, Wenjie Li, and Hui Xiong. 2024.
\newblock Jobformer: Skill-aware job recommendation with semantic-enhanced transformer.
\newblock \emph{ACM Transactions on Knowledge Discovery from Data}.

\bibitem[{Guenole and Feinzig(2018)}]{guenole2018business}
Nigel Guenole and Sheri Feinzig. 2018.
\newblock The business case for ai in hr.
\newblock \emph{With Insights and Tips on Getting Started. Armonk: IBM Smarter Workforce Institute, IBM Corporation}.

\bibitem[{Jain(2023)}]{tiktoken}
Shantanu Jain. 2023.
\newblock \href {https://github.com/openai/tiktoken} {tiktoken}.

\bibitem[{Kothari et~al.(2024)Kothari, Mehta, Patil, and Hole}]{kothari2024interviewease}
Param Kothari, Paras Mehta, Srushti Patil, and Varsha Hole. 2024.
\newblock Interviewease: Ai-powered interview assistance.

\bibitem[{Lee et~al.(2019)Lee, Jagannath, Aggarwal, Sridar, Wilde, Hill, and Chen}]{lee2019intelligent}
Terri Lee, Krithika Jagannath, Nitin Aggarwal, Ramamurti Sridar, Shawn Wilde, Timothy Hill, and Yu~Chen. 2019.
\newblock Intelligent career advisers in your pocket? a need assessment study of chatbots for student career advising.

\bibitem[{Li et~al.(2021)Li, Yuan, Kamarthi, Moghaddam, and Jin}]{li2021data}
Guoyan Li, Chenxi Yuan, Sagar Kamarthi, Mohsen Moghaddam, and Xiaoning Jin. 2021.
\newblock Data science skills and domain knowledge requirements in the manufacturing industry: A gap analysis.
\newblock \emph{Journal of Manufacturing Systems}, 60:692--706.

\bibitem[{{MongoDB}()}]{MongoDB}
{MongoDB}.
\newblock Mongodb: Built by developers, for developers.
\newblock \url{https://www.mongodb.com/}.

\bibitem[{Montgomery(2017)}]{montgomery2017mapping}
Beronda~L Montgomery. 2017.
\newblock Mapping a mentoring roadmap and developing a supportive network for strategic career advancement.
\newblock \emph{Sage Open}, 7(2):2158244017710288.

\bibitem[{OpenAI(2022)}]{ChatGPT}
OpenAI. 2022.
\newblock Chatgpt: An ai language model.
\newblock \url{https://www.openai.com}.

\bibitem[{{Qdrant}()}]{qdrant}
{Qdrant}.
\newblock High-performance vector search at scale.
\newblock \url{https://qdrant.tech/}.

\bibitem[{Ramírez(2018)}]{2018fastapi}
Sebastián Ramírez. 2018.
\newblock \href {https://fastapi.tiangolo.com/} {Fastapi}.

\bibitem[{Rao et~al.(2024)Rao, Lipizzi, and Mansouri}]{rao2024anatomy}
Balaji Rao, Carlo Lipizzi, and Mo~Mansouri. 2024.
\newblock Anatomy of an ai economy.
\newblock In \emph{2024 IEEE International Symposium on Systems Engineering (ISSE)}, pages 1--8. IEEE.

\bibitem[{Shilaskar et~al.(2024)Shilaskar, Bhatlawande, Sawle, Gupta, and Buche}]{shilaskar2024conversational}
Swati Shilaskar, Shripad Bhatlawande, Prathamesh Sawle, Sanchita Gupta, and Reva Buche. 2024.
\newblock Conversational ai for career counseling.
\newblock In \emph{2024 MIT Art, Design and Technology School of Computing International Conference (MITADTSoCiCon)}, pages 1--5. IEEE.

\bibitem[{Sinha et~al.(2021)Sinha, Amir Khusru~Akhtar, and Kumar}]{sinha2021resume}
Arvind~Kumar Sinha, Md~Amir Khusru~Akhtar, and Ashwani Kumar. 2021.
\newblock Resume screening using natural language processing and machine learning: A systematic review.
\newblock \emph{Machine Learning and Information Processing: Proceedings of ICMLIP 2020}, pages 207--214.

\bibitem[{Xu et~al.(2018)Xu, Yu, Guo, Teng, and Xiong}]{xu2018extracting}
Huang Xu, Zhiwen Yu, Bin Guo, Mingfei Teng, and Hui Xiong. 2018.
\newblock Extracting job title hierarchy from career trajectories: A bayesian perspective.
\newblock In \emph{IJCAI}, pages 3599--3605.

\end{thebibliography}
\bibliographystyle{acl_natbib}

\newpage
\textbf{Appendix}
\newline
\appendix \textbf{A Limitations}\\
The current iteration of Steve heavily relies on manually curated JSON career paths, skill requirements, and benchmarks. This reliance influences the generalizability of the system in cases where career progression and skills are more fluid or cannot be captured in traditional terms.

The proficiency levels are determined using implicit factors such as resume mentions and Q\&A responses, which could lead to over or underestimation. Moreover, our system lacks a concrete form of evaluation to attest to the candidate's proficiency in the skills mentioned.

Although there is a tracker, we do not have explicit mechanisms for collecting feedback to refine and validate its recommendations. Additionally, the system does not account for preferences such as budget, learning style, or time commitment, which could significantly impact its utility for the candidate.There are simply some of issues we have identified. We will continur to build on this iteration through rigirous feedback and crowd debugging, both technical and non-technical.\newline

\appendix \textbf{B Future Work: Some Ideas}\\
For the next iteration, we are experimenting with evaluation mechanisms to empirically assess skill proficiency, as well as automate the creation and updating of career trees using real-time job market data and labor market analytics. This would help in generating dynamic and up-to-date recommendations.

We are also working on building out functionalities to provide detailed visualizations of career progression options and skill gaps for better user understanding. Steve would also benefit from inputs from SMEs across domains, and having multiple organizations adopt our system would help in both developing skills and finding roles that require these skills within various industries.

\end{document}